# Gilbert and Landau-Lifshitz damping in the presense of spin-torque

Neil Smith
San Jose Reserch Center, Hitachi Global Strage Technologies, San Jose, CA 95135

A recent article by Stiles *et al.* (cond-mat/0702020) argued in favor of the Landau-Lifshitz damping term in the micromagnetic equations of motion over that of the more commonly accepted Gilbert damping form. Much of their argument revolved around spin-torque driven domain wall motion in narrow magnetic wires, since the presence of spin-torques can more acutely draw a distinction between the two forms of damping. In this article, the author uses simple arguments and examples to offer an alternative point of view favoring Gilbert.

## I. PRELIMINARIES

The Gilbert[1] (G) or Landau-Lifshitz[2] (LL) equations of motion for unit magnetization vector $\hat{\boldsymbol{m}}(\boldsymbol{r},t) \equiv \boldsymbol{M}(\boldsymbol{r},t)/M_s$ are formally described by the generic form

$$\begin{aligned} d\hat{\boldsymbol{m}}/dt &= -\gamma[\hat{\boldsymbol{m}} \times (\boldsymbol{H}_{\text{tot}} + \boldsymbol{H}_{\text{damp}})] \\ \boldsymbol{H}_{\text{tot}} &\equiv \boldsymbol{H}_{\text{eff}} + \boldsymbol{H}_{\text{NC}} \\ \boldsymbol{H}_{\text{eff}} &\equiv -1/M_s\, \partial E/\partial\hat{\boldsymbol{m}} \ \text{(variational derivative)} \end{aligned} \tag{1}$$

where $M_s$ the saturation magnetization, and $\gamma$ is the gyromagnetic ratio (taken here to be a positive constant). The total (physical) field by has contributions from the usual "effective field" term $\boldsymbol{H}_{\text{eff}}$, plus that of a "nonconservative-field" $\boldsymbol{H}_{\text{NC}}$ that is supposed *not* to be derivable from the $\hat{\boldsymbol{m}}$-gradient of the (internal) free-energy density functional $E(\hat{\boldsymbol{m}})$. Although also nonconservative by definition, the "damping-field" $\boldsymbol{H}_{\text{damp}}$ is primarily a mathematical vehicle for describing a physical damping torque $M_s\hat{\boldsymbol{m}} \times \boldsymbol{H}_{\text{damp}}$, and is properly treated separately. For most of the remainder of this article, any spatial dependence of $\hat{\boldsymbol{m}}(\boldsymbol{r},t)$ will be implicitly understood.

As was described by Brown,[3] the Gilbert equations of motion may be derived using standard techniques of Lagrangian mechanics.[4] In particular, a phenomenological damping of the motion in included via the use of a Rayleigh dissipation function $\Re(d\hat{\boldsymbol{m}}/dt)$:

$$\begin{aligned} \Re &= (\alpha_G/2\gamma)\, M_s \left|d\hat{\boldsymbol{m}}/dt\right|^2 \\ \boldsymbol{H}^{\text{G}}_{\text{damp}} &\equiv -1/M_s\, \partial\Re/\partial(d\hat{\boldsymbol{m}}/dt) = -(\alpha_G/\gamma)\, d\hat{\boldsymbol{m}}/dt \end{aligned} \tag{2}$$

where dimensionless $\alpha_{\text{G}}$ is the Gilbert damping parameter. By definition,[4] $2\Re = -\boldsymbol{H}^{\text{G}}_{\text{damp}} \cdot d\hat{\boldsymbol{m}}/dt = \alpha_G M_s/\gamma\ \left|d\hat{\boldsymbol{m}}/dt\right|^2$ is the instantaneous rate of energy lost from the magnetization system to its thermal environment (e.g., to the lattice) due to the *viscous* "friction" represented by the damping field $\boldsymbol{H}^{\text{G}}_{\text{damp}} \propto -d\hat{\boldsymbol{m}}/dt$.

The Lagrangian method is well suited to include nonconservative fields $\boldsymbol{H}_{\text{NC}} \neq 0$, which can be generally defined using the principles of virtual work:[3,4]

$$\begin{aligned} \delta W_{\text{NC}} &= M_s\, \boldsymbol{H}_{\text{NC}} \cdot \delta\hat{\boldsymbol{m}} = M_s\, (\hat{\boldsymbol{m}} \times \boldsymbol{H}_{\text{NC}}) \cdot \delta\boldsymbol{\theta} \\ \Rightarrow N &= M_s (\hat{\boldsymbol{m}} \times \boldsymbol{H}_{\text{NC}}) \Leftrightarrow \boldsymbol{H}_{\text{NC}} = 1/M_s\, (\boldsymbol{N} \times \hat{\boldsymbol{m}}) \end{aligned} \tag{3}$$

The latter expression is useful in cases (e.g., spin-torques) where the torque density functional $\boldsymbol{N}(\hat{\boldsymbol{m}})$ is specified. Treating $M_s$ as fixed, the (virtual) displacement $\delta\hat{\boldsymbol{m}}$ is of the form $\delta\hat{\boldsymbol{m}} = \delta\boldsymbol{\theta} \times \hat{\boldsymbol{m}}$, and only the orthogonal components of the torque $\boldsymbol{N} \leftrightarrow \hat{\boldsymbol{m}} \times \boldsymbol{N} \times \hat{\boldsymbol{m}}$ are physically significant.

Combining (1) and (2) gives the Gilbert equations:

$$d\hat{\boldsymbol{m}}/dt = -\gamma\,(\hat{\boldsymbol{m}} \times \boldsymbol{H}_{\text{tot}}) + \alpha_G\,(\hat{\boldsymbol{m}} \times d\hat{\boldsymbol{m}}/dt) \tag{4}$$

As is well known, the G equations of (4) may be rearranged into their equivalent (and perhaps more common) form:

$$d\hat{\boldsymbol{m}}/dt = \frac{-\gamma}{1+\alpha_{\text{G}}^2}[\hat{\boldsymbol{m}} \times \boldsymbol{H}_{\text{tot}} + \alpha_G\, \hat{\boldsymbol{m}} \times (\hat{\boldsymbol{m}} \times \boldsymbol{H}_{\text{tot}})] \tag{5}$$

With regard to the LL equations, the form of $\boldsymbol{H}^{\text{LL}}_{\text{damp}}$ is not uniquely defined in problems where $\boldsymbol{H}_{\text{NC}} \neq 0$, which have only come to the forefront with the recent interest in spin-torque phenomena. Two definitions considered are

$$\boldsymbol{H}^{\text{LL}}_{\text{damp}} \equiv \alpha_{\text{LL}}\,(\hat{\boldsymbol{m}} \times \boldsymbol{H}_{\text{eff}}), \tag{6a}$$

$$\boldsymbol{H}^{\text{LL}}_{\text{damp}} \equiv \alpha_{\text{LL}}\,(\hat{\boldsymbol{m}} \times \boldsymbol{H}_{\text{tot}}) \tag{6b}$$

The first definition of (6a) is the historical/conventional form of LL, and is that employed by Stiles *et al.*[5] However, in this author's view, there is no *a-priori* reason, other than historical, to *not* replace $\boldsymbol{H}_{\text{eff}} \to \boldsymbol{H}_{\text{tot}}$ as in (6b). Doing so yields a form of LL that retains it "usual" equivalence (i.e., to first order in $\alpha$) to G whether or not $\boldsymbol{H}_{\text{NC}} \neq 0$, as is seen by comparing (5) and (6b). The form of (6b) treats both $\boldsymbol{H}_{\text{eff}}$ and $\boldsymbol{H}_{\text{NC}}$ on an equal footing.

Nonetheless, to facilitate a comparative discussion with the analysis of Stiles *et al.*,[5] (6a) will henceforth be used to define what will be referred to below as the LL equations of motion:

$$d\hat{\boldsymbol{m}}/dt = -\gamma[\hat{\boldsymbol{m}}\times\boldsymbol{H}_{\text{tot}} + \alpha_{\text{LL}}\,\hat{\boldsymbol{m}}\times(\hat{\boldsymbol{m}}\times\boldsymbol{H}_{\text{eff}})] \qquad (7)$$

In cases of present interest where $\boldsymbol{H}_{\text{NC}} \neq 0$, the difference between G in (4) (or (5)) and the form of LL given in (7) are *first* order in the damping parameter, and thus of a more fundamental nature. These differences are the subject of the remainder of this article.

## II. SPIN-TORQUE EXAMPLES

Two distinct situations where spin-torque effects have garnered substantial interest are those of CPP-GMR nanopillars, and spin-torque driven domain wall motion in nanowires as was considered in Ref. 5. The spin-torque function $\boldsymbol{N}_{\text{ST}}(\hat{\boldsymbol{m}}) = \boldsymbol{N}_{\text{ad}}(\hat{\boldsymbol{m}}) + \boldsymbol{N}_{\text{nad}}(\hat{\boldsymbol{m}})$ is taken to have a predominant "adiabatic" component $\boldsymbol{N}_{\text{ad}}(\hat{\boldsymbol{m}})$, along with a small "nonadiabatic" component $\boldsymbol{N}_{\text{nad}}(\hat{\boldsymbol{m}})$ described phenomenologically by the relation $\boldsymbol{N}_{\text{nad}} \equiv -\beta\,\hat{\boldsymbol{m}}\times\boldsymbol{N}_{\text{ad}}$, with $\beta \ll 1$. In the case of a narrow nanowire along the $\hat{\boldsymbol{x}}$-axis, with magnetization $\hat{\boldsymbol{m}}(x)$ and electron current density $\boldsymbol{J}_e = J_e\hat{\boldsymbol{x}}$, the torque function $\boldsymbol{N}_{\text{ST}}(\hat{\boldsymbol{m}})$ and associated field $\boldsymbol{H}_{\text{ST}}(\hat{\boldsymbol{m}})$ (see (2)) are described by[5]

$$\begin{aligned}\boldsymbol{N}_{\text{ad}}(\hat{\boldsymbol{m}}) &= (\hbar P J_e/2e)\,(d\hat{\boldsymbol{m}}/dx)\\ \boldsymbol{H}_{\text{ST}} &= -(\hbar P J_e/2M_s e)\,(\hat{\boldsymbol{m}}\times d\hat{\boldsymbol{m}}/dx + \beta\, d\hat{\boldsymbol{m}}/dx)\end{aligned} \qquad (8)$$

where *P* is the spin-polarization of the electron current.

To check if $\boldsymbol{H}_{\text{ST}}$ is conservative, one can "discretize" the spatial derivatives appearing in (8) in the form $d\hat{\boldsymbol{m}}/dx\big|_{x=x_i} \to (\hat{\boldsymbol{m}}_{i+1} - \hat{\boldsymbol{m}}_{i-1})/2\Delta x$, where $\hat{\boldsymbol{m}}_i \equiv \hat{\boldsymbol{m}}(x_i)$ and $\Delta x \equiv x_{i+1} - x_i$, not unlike the common micromagnetics approximation. For a conservative $\boldsymbol{H}$-field where $\boldsymbol{H}_i \propto \partial E/\partial \boldsymbol{m}_i$, the set of $3\times 3$ Cartesian tensors[6] $\overleftrightarrow{H}^{uv}_{ij} \equiv \partial \boldsymbol{H}_i/\partial \boldsymbol{m}_j \propto \partial^2 E/\partial \boldsymbol{m}_i \partial \boldsymbol{m}_j$ will be symmetric, i.e., $\overleftrightarrow{H}^{uv}_{ij} = \overleftrightarrow{H}^{vu}_{ji}$, under simultaneous reversal of spatial indices $i, j$ and vector indices $u, v = x, y,$ or $z$. For the adiabatic term in (6), it can be readily shown that the $\overleftrightarrow{H}^{uv}_{ij}$ are in general *asymmetric*, i.e., always antisymmetric in vector indices (due to cross product), but asymmetric in spatial indices $i, j = i \pm 1$, being antisymmetric here only for locally uniform magnetization $\hat{\boldsymbol{m}}_{i\pm1} = \hat{\boldsymbol{m}}_i$. The nonadiabatic term yields an $\hat{\boldsymbol{m}}$-independent $\overleftrightarrow{H}^{uv}_{ij}$ that is always antisymmetric, i.e., symmetric in vector indices, but antisymmetric in spatial indices $i, j = i \pm 1$. The conclusion here that $\boldsymbol{H}_{\text{ST}}$ is in general nonconservative agrees with that found in Ref. 5, by way of a rather different argument.

Another well known example is a nanopillar stack with only two ferromagnetic (FM) layers, the "reference" layer having a magnetization $\hat{\boldsymbol{m}}_{\text{ref}}$ rigidly fixed in time, and a dynamically variable "free" layer $\hat{\boldsymbol{m}}_{\text{free}}(t) = \hat{\boldsymbol{m}}(t)$. As described by Slonczewski,[7] the (adiabatic) spin-torque density function and field $\boldsymbol{H}_{\text{ST}}(\hat{\boldsymbol{m}})$ is given by:

$$\begin{aligned}\boldsymbol{N}_{\text{ad}} &= -g(\hat{\boldsymbol{m}}_{\text{ref}}\cdot\hat{\boldsymbol{m}})\,(\hbar P J_e/4e\,t_{\text{free}})[\hat{\boldsymbol{m}}\times\hat{\boldsymbol{m}}_{\text{ref}}\times\hat{\boldsymbol{m}}]\\ \boldsymbol{H}_{\text{ST}} &= -g(\hat{\boldsymbol{m}}_{\text{ref}}\cdot\hat{\boldsymbol{m}})\,(\hbar P J_e/4e\,M_s t_{\text{free}})\\ &\qquad\times[(\hat{\boldsymbol{m}}_{\text{ref}}\times\hat{\boldsymbol{m}}) + \beta\,\hat{\boldsymbol{m}}_{\text{ref}}]\end{aligned} \qquad (9)$$

where $t_{\text{free}}$ is the free layer thickness, and $g(\hat{\boldsymbol{m}}_{\text{ref}}\cdot\hat{\boldsymbol{m}})$ is a function of order unity, the details of which are not relevant to the present discussion. From the the $\overleftrightarrow{H}^{uv}$-tensor, or by simple inspection, the adiabatic $\hat{\boldsymbol{m}}_{\text{ref}}\times\hat{\boldsymbol{m}}$ term in (7) is manifestly nonconservative. However, approximating $g(\hat{\boldsymbol{m}}_{\text{ref}}\cdot\hat{\boldsymbol{m}}) \sim$ constant, the conservative nonadiabatic term resembles a magnetic field described by the $\hat{\boldsymbol{m}}$-gradient of an Zeeman-like energy function $E_{\text{nad}} \propto \hat{\boldsymbol{m}}_{\text{ref}}\cdot\hat{\boldsymbol{m}}$. The remaining discussion will restrict attention to nonconservative contributions.

## III. STATIONARY SOLUTIONS OF G AND LL

With $\boldsymbol{H}_{\text{NC}} \to \boldsymbol{H}_{\text{ST}}$, stationary (i.e, $d\hat{\boldsymbol{m}}/dt = 0$) solutions $\hat{\boldsymbol{m}}_0$ of G-equations (4) satisfy the conditions that

$$\begin{aligned}&\hat{\boldsymbol{m}}_0\times(\boldsymbol{H}_{\text{eff}} + \boldsymbol{H}_{\text{ST}}) = 0;\ \boldsymbol{H}^{\text{G}}_{\text{damp}} \propto d\hat{\boldsymbol{m}}/dt = 0\\ &\hat{\boldsymbol{m}}_0\times\boldsymbol{H}_{\text{ST}} \neq 0 \Rightarrow \hat{\boldsymbol{m}}_0\times\boldsymbol{H}_{\text{eff}} = -\hat{\boldsymbol{m}}_0\times\boldsymbol{H}_{\text{ST}}\end{aligned} \qquad (10)$$

The clear and physically intuitive interpretation of (10) is that stationary state $\hat{\boldsymbol{m}}_0$ satisfies a condition of zero physical torque, $\hat{\boldsymbol{m}}_0\times\boldsymbol{H}_{\text{tot}} = 0$, including both conservative ($\boldsymbol{H}_{\text{eff}}$) and nonconservative spin-torque ($\boldsymbol{H}_{\text{ST}}$) fields. Being viscous in nature, the G damping torque $\hat{\boldsymbol{m}}_0\times\boldsymbol{H}^{\text{G}}_{\text{damp}} \propto d\hat{\boldsymbol{m}}/dt \equiv 0$ independently vanishes..

Previous measurements[6] of the angular dependence of spin-torque critical currents $J^{\text{crit}}_e(\hat{\boldsymbol{m}}_{\text{ref}}\cdot\hat{\boldsymbol{m}})$ in CPP-GMR nanopillar systems by this author and colleagues demonstrated the existence of such stationary states with *non*-collinear $\hat{\boldsymbol{m}}_{\text{ref}}\times\hat{\boldsymbol{m}}_0 \neq 0$ and $0 < |J_e| < J^{\text{crit}}_e$. In this situation, it follows from (9) and (10) that the stationary state $\hat{\boldsymbol{m}}_0$ satisfies $-\hat{\boldsymbol{m}}_0\times\boldsymbol{H}_{\text{ST}} = \hat{\boldsymbol{m}}_0\times\boldsymbol{H}_{\text{eff}} \neq 0$. It is noted that the last result implies that $\hat{\boldsymbol{m}}_0$ is *not* a (thermal) equilibrium state which minimizes the free energy $E(\hat{\boldsymbol{m}})$, i.e., $\delta E / \delta \hat{\boldsymbol{m}} = (\partial E / \partial \hat{\boldsymbol{m}}) \cdot (\delta\theta \times \hat{\boldsymbol{m}}_0) \propto (\hat{\boldsymbol{m}}_0 \times \boldsymbol{H}_{\text{eff}}) \cdot \delta\theta \neq 0$ for arbitrary $\delta\theta$.

In the present described circumstance of stationary $\hat{\boldsymbol{m}}_0$ with $\hat{\boldsymbol{m}}_0 \times \boldsymbol{H}_{\text{ST}} \neq 0$, the LL equations of (7) differ from G in a fundamental respect. Setting $d\hat{\boldsymbol{m}} / dt = 0$ in (7) yields

$$\hat{\boldsymbol{m}}_0 \times (\boldsymbol{H}_{\text{eff}} + \boldsymbol{H}_{\text{ST}}) = -\alpha_{\text{LL}}\, \hat{\boldsymbol{m}}_0 \times (\hat{\boldsymbol{m}}_0 \times \boldsymbol{H}_{\text{eff}}) \qquad (11)$$

Like (10), (11) implies that $\hat{\boldsymbol{m}}_0 \times \boldsymbol{H}_{\text{eff}} \neq 0$ when $\hat{\boldsymbol{m}}_0 \times \boldsymbol{H}_{\text{ST}} \neq 0$. However, (11) also imply a *static, nonzero* physical torque $\hat{\boldsymbol{m}}_0 \times \boldsymbol{H}_{\text{tot}} \neq 0$, along with a *static, nonzero damping torque* $\hat{\boldsymbol{m}}_0 \times \boldsymbol{H}^{\text{LL}}_{\text{damp}} \neq 0$ (see (6a)) to cancel it out. In simple mechanical terms, the latter amounts to non-viscous "static-friction". It has no analogue with G in any circumstance, or with LL in conventional situations with $\boldsymbol{H}_{\text{NC}} \leftrightarrow \boldsymbol{H}_{\text{ST}} = 0$ and $\hat{\boldsymbol{m}}_0 \leftrightarrow$ equilibrium for which LL damping was originally developed as a phenomenological damping form. It further contradicts the viscous (or $d\hat{\boldsymbol{m}}/dt$-dependent) nature of the damping mechanisms described by physical (rather than phenomenological) based theoretical models[8,9].

The above arguments ignored thermal fluctuations of $\hat{\boldsymbol{m}}$. However, thermal fluctuations[10] scale approximately as $kT / (\boldsymbol{m}_0 \cdot \boldsymbol{H}_{\text{eff}})^2$, while (10) or (11) are scale-invariant with $|\boldsymbol{H}|$. In the simple CPP nanopillar example of (9), one can (conceptually at least) continually increase both $|J_e|$ and an applied field $\boldsymbol{H}_{\text{app}}$ contribution to $\boldsymbol{H}_{\text{eff}}$ to scale up $|\hat{\boldsymbol{m}}_0 \times \boldsymbol{H}_{\text{ST}}|$ and $|\hat{\boldsymbol{m}}_0 \times \boldsymbol{H}_{\text{eff}}|$ while approximately keeping a fixed stationary state $\hat{\boldsymbol{m}}_0$ (satisfying $\hat{\boldsymbol{m}}_0 \times \hat{\boldsymbol{m}}_{\text{ref}} \neq 0$ with fixed $\hat{\boldsymbol{m}}_{\text{ref}}$). However, unique to LL equations (11) based on (6a) is the additional requirement that the static damping mechanism be able to produce an $|\boldsymbol{H}^{\text{LL}}_{\text{damp}}| \propto |\hat{\boldsymbol{m}} \times \boldsymbol{H}_{\text{eff}}|$ which similarly scales (without limit). This author finds this a physically unreasonable proposition.

## IV. ENERGY ACCOUNTING

If one ignores/forgets the Lagrangian formulation[3] of the Gilbert equations (4), one may derive the following energy relationships, substituting the right side of (4) for evaluating vector products of form $\boldsymbol{H} \cdot d\hat{\boldsymbol{m}} / dt$:

$$\begin{aligned} 1/M_s\, (dE/dt = \partial E / \partial \hat{\boldsymbol{m}} \cdot d\hat{\boldsymbol{m}}/dt) &\equiv -\boldsymbol{H}_{\text{eff}} \cdot d\hat{\boldsymbol{m}}/dt \\ &= \gamma \boldsymbol{H}_{\text{eff}} \cdot (\hat{\boldsymbol{m}} \times \boldsymbol{H}_{\text{NC}}) - \alpha\, \boldsymbol{H}_{\text{eff}} \cdot (\hat{\boldsymbol{m}} \times d\hat{\boldsymbol{m}}/dt) \\ &= -\gamma \boldsymbol{H}_{\text{NC}} \cdot (\hat{\boldsymbol{m}} \times \boldsymbol{H}_{\text{eff}}) - \alpha\, \boldsymbol{H}_{\text{eff}} \cdot (\hat{\boldsymbol{m}} \times d\hat{\boldsymbol{m}}/dt) \end{aligned} \qquad (12a)$$

$$\begin{aligned} 1/M_s\, dW_{\text{NC}}/dt &\equiv \boldsymbol{H}_{\text{NC}} \cdot d\hat{\boldsymbol{m}}/dt \\ &= -\gamma \boldsymbol{H}_{\text{NC}} \cdot (\hat{\boldsymbol{m}} \times \boldsymbol{H}_{\text{eff}}) + \alpha\, \boldsymbol{H}_{\text{NC}} \cdot (\hat{\boldsymbol{m}} \times d\hat{\boldsymbol{m}}/dt) \end{aligned} \qquad (12b)$$

$$\begin{aligned} |d\hat{\boldsymbol{m}}/dt|^2 &= d\hat{\boldsymbol{m}}/dt \cdot (-\gamma\, \hat{\boldsymbol{m}} \times \boldsymbol{H}_{\text{tot}}) \\ &= \gamma\, (\boldsymbol{H}_{\text{eff}} + \boldsymbol{H}_{\text{NC}}) \cdot (\hat{\boldsymbol{m}} \times d\hat{\boldsymbol{m}}/dt) \end{aligned} \qquad (12c)$$

Subtracting (12b) from (12a), and using (12c) one finds

$$\begin{aligned} \text{G}: \ dE/dt &= dW_{\text{NC}}/dt \ - \ \alpha_{\text{G}}\, M_s/\gamma\, |d\hat{\boldsymbol{m}}/dt|^2 \\ &= dW_{\text{NC}}/dt \ + \ M_s\, \boldsymbol{H}^{\text{G}}_{\text{damp}} \cdot d\hat{\boldsymbol{m}}/dt \end{aligned} \qquad (13)$$

The result of (13) is essentially a statement of energy conservation. Namely, that the rate of change of the internal free energy (density) of the magnetic system is give by the work done on the system by the (external) nonconservative forces/fields $\boldsymbol{H}_{\text{NC}}$, minus the loss of energy (to the lattice) due to damping. The G damping term in (13) is (not surprisingly) the same as expected from (2). It is a *strictly lossy, negative-definite* contribution to $dE/dt$.

Over a finite interval of motion from time $t_1$ to $t_2$, the change $\Delta E = E(\hat{\boldsymbol{m}}_2) - E(\hat{\boldsymbol{m}}_1)$ is, from (12b) and (13):

$$\Delta E = M_s \int_{t_1}^{t_2} dt\, (\boldsymbol{H}_{\text{NC}}(\hat{\boldsymbol{m}}) - \alpha_{\text{G}}/\gamma\ d\hat{\boldsymbol{m}}/dt) \cdot \frac{d\hat{\boldsymbol{m}}}{dt} \qquad (14)$$

Since $\boldsymbol{H}_{\text{NC}}$ is nonconservative, the work $\Delta W_{\text{NC}}$ is path-dependent, and so use of (14) requires independent knowledge of the solution $\hat{\boldsymbol{m}}(t_1 \leq t \leq t_2)$ of (4). Since $\hat{\boldsymbol{m}}(t)$ itself depends on $\alpha_G$, the $\boldsymbol{H}_{\text{NC}}$ term's contribution to (14) also can vary with $\alpha_G$. Regardless, $\Delta E > 0$ can *only* result in the case of a *positive* amount of work $\Delta W_{\text{NC}} = M_s \int_{t_1}^{t_2} (\boldsymbol{H}_{\text{NC}} \cdot d\hat{\boldsymbol{m}}/dt)\, dt$ done by $\boldsymbol{H}_{\text{NC}}$.

Working out the results analogous to (12a,b) for the LL equations of (6a) and (7), one finds

$$\begin{aligned} \text{LL}: \ dE/dt &= dW_{\text{NC}}/dt \ + \ M_s\, \boldsymbol{H}^{\text{LL}}_{\text{damp}} \cdot d\hat{\boldsymbol{m}}/dt \\ \boldsymbol{H}^{\text{LL}}_{\text{damp}} \cdot d\hat{\boldsymbol{m}}/dt &= -\alpha\gamma\, (\hat{\boldsymbol{m}} \times \boldsymbol{H}_{\text{eff}}) \cdot (\hat{\boldsymbol{m}} \times \boldsymbol{H}_{\text{tot}}) \end{aligned} \qquad (15)$$

The form of (15) is the same as the latter result in (13). However, unlike G, the LL damping term in (15) is *not* manifestly negative-definite, except when $\boldsymbol{H}_{\text{NC}} = 0$.

The results of (13)-(15) apply equally to situations where one integrates over the spatial distribution of $\hat{\boldsymbol{m}}(\boldsymbol{r}, t)$ to evaluate the total system free energy, rather than (local) free energy density. Total time derivatives $d/dt$ may be replaced by partial derivatives $\partial / \partial t$ where appropriate.

Dropping terms of order $\alpha_{LL}^2$ (and simplifying notation $\alpha_{LL} \to \alpha$), (7) is easily transformed to a Gilbert-like form:

$$\text{LL}: \frac{d\hat{\boldsymbol{m}}}{dt} = -\gamma\,\hat{\boldsymbol{m}} \times [\boldsymbol{H}_{\text{tot}} - \alpha\,(\hat{\boldsymbol{m}} \times \boldsymbol{H}_{\text{NC}})] + \alpha\,(\hat{\boldsymbol{m}} \times \frac{d\hat{\boldsymbol{m}}}{dt}) \quad (16)$$

which differs from G in (4) by the term $\alpha\,(\hat{\boldsymbol{m}} \times \boldsymbol{H}_{\text{NC}})$ which is first order in both $\alpha$ and $\boldsymbol{H}_{\text{NC}}$. For the "wire problem" described by (8), the equations of motion become

$$\begin{aligned} \text{G}: \frac{d\hat{\boldsymbol{m}}}{dt} + v\frac{d\hat{\boldsymbol{m}}}{dx} &= -\gamma\,\hat{\boldsymbol{m}} \times \boldsymbol{H}_{\text{eff}} + \alpha\,\hat{\boldsymbol{m}} \times (\frac{d\hat{\boldsymbol{m}}}{dt} + v\frac{\beta}{\alpha}\frac{d\hat{\boldsymbol{m}}}{dx}) \\ \text{LL}: \frac{d\hat{\boldsymbol{m}}}{dt} + v\frac{d\hat{\boldsymbol{m}}}{dx} &= -\gamma\,\hat{\boldsymbol{m}} \times \boldsymbol{H}_{\text{eff}} + \alpha\,\hat{\boldsymbol{m}} \times (\frac{d\hat{\boldsymbol{m}}}{dt} + v\frac{\alpha+\beta}{\alpha}\frac{d\hat{\boldsymbol{m}}}{dx}) \end{aligned} \quad (17)$$

where $v = \hbar\gamma P J_e / 2M_s e$, and terms of order $\alpha\beta$ are dropped for LL. As noted previously,[5,9,11] (17) permits "translational" solutions $\hat{\boldsymbol{m}}(x,t) = \hat{\boldsymbol{m}}_{\text{eq}}(x - vt)$ when $\beta = \alpha$ (G) or $\beta = 0$ (LL), with $\hat{\boldsymbol{m}}_{\text{eq}}(x)$ the *static*, equilibrium (minimum $E$) solution of $(\hat{\boldsymbol{m}}_{\text{eq}} \times \boldsymbol{H}_{\text{eff}}) = 0$. Evaluating $dW_{\text{ST}}/dt = M_s\,\boldsymbol{H}_{\text{ST}} \cdot d\hat{\boldsymbol{m}}/dt$ by taking $\boldsymbol{H}_{\text{ST}}$ from (8), and $d\hat{\boldsymbol{m}}/dt \to -v\,\hat{\boldsymbol{m}}'_{\text{eq}}(x - vt)$ with $\hat{\boldsymbol{m}}'(q) \equiv d\hat{\boldsymbol{m}}/dq$, one finds that $dW_{\text{ST}}/dt = (\beta v^2 M_s/\gamma)\left|\hat{\boldsymbol{m}}'_{\text{eq}}(x - vt)\right|^2$. In translational cases where $d\hat{\boldsymbol{m}}/dt$ is *exactly collinear* to $d\hat{\boldsymbol{m}}/dx$, only the nonadiabatic term does work on the $\hat{\boldsymbol{m}}$-system.

Interestingly, the energy interpretation of these translational solutions is *very* different for G or LL. For G, the positive rate of work $dW_{\text{ST}}/dt$ when $\beta = \alpha$ exactly balances the negative damping loss as given in (13), the latter always nonzero and scaling as $v^2$. For LL by contrast, the work done by $\boldsymbol{H}_{\text{ST}}$ vanishes when $\beta = 0$, matching the damping loss which, from (6a) or (15), is *always zero* since $(\hat{\boldsymbol{m}}_{\text{eq}} \times \boldsymbol{H}_{\text{eff}}) = 0$ *regardless* of $v$. If $\hat{\boldsymbol{m}}_{\text{eq}}(x)$ is a sharp domain wall, $d\hat{\boldsymbol{m}}/dt = -v\,\hat{\boldsymbol{m}}'_{\text{eq}}(x - vt)$ represents, from a spatially *local* perspective at a *fixed* point $x$, an abrupt, irreversible, non-equilibrium reorientation of $\hat{\boldsymbol{m}}$ at/near time $t \approx x/v$ when the wall core passes by. The prediction of LL/(6a) that this magnetization reversal could take place *locally* (at arbitrarily large $v$), with the complete absence of the spin-orbit coupled, electron scattering processes[8] that lead to spin-lattice damping/relaxation in all other known circumstances (e.g., external field-driven domain wall motion) is, in the view of this author, a rather dubious, nonphysical aspect of (6a) when $\boldsymbol{H}_{\text{ST}} \neq 0$.

Stiles et al.[5] report that micromagnetic computations using G in the case $\beta = 0$ show (non-translational) time/distance limited domain wall displacement, resulting in a net *positive* increase $\Delta E$. They claim that 1) "spin transfer torques do not change the energy of the system", and that 2) "Gilbert damping torque is the only torque that changes the energy". Accepting $\Delta E > 0$ as accurate, it is this author's view that the elementary physics/mathematics leading to (13) and (14) demonstrably prove that both of these claims must be incorrect (error in the first perhaps leading to the misinterpretation of the second). On a related point, the results of (13) and (15) shows that *excluding* work $dW_{\text{ST}}/dt$ or $\Delta W_{\text{ST}}$, only LL-damping may possibly lead to a positive contribution to $dE/dt$ or $\Delta E$ when $\boldsymbol{H}_{\text{ST}} \neq 0$, in apparent contradiction to the claim in Ref. 5 that LL damping "uniquely and irreversibly reduces magnetic free energy when spin-transfer torque is present".

## ACKNOWLEDGMENTS

The author would like to acknowledge email discussions on these or related topics with W. Saslow and R. Duine, as well as an extended series of friendly discussions with Mark Stiles. Obviously, the latter have not (as of yet) achieved a mutually agreed viewpoint on this subject.

## REFERENCES

[1] T. L. Gilbert, Armour Research Report, May 1956; IEEE Trans. Magn., **40**, 3343 (2004).
[2] L. Landau and E. Lifshitz, Phys. Z. Sowjet **8**, 153 (1935).
[3] W. F. Brown, *Micromagnetics* (Krieger, New York 1978).
[4] H. Goldstein, *Classical Mechanics*, (Addison Wesley, Reading Massachusetts, 1950).
[5] M. D. Stiles, W. M. Saslow, M. J. Donahue, and A. Zangwill, arXiv:cond-mat/0702020.
[6] N. Smith, J. A. Katine, J. R. Childress, and M. J. Carey, IEEE Trans. Magn. **41**, 2935 (2005); N. Smith, J Appl. Phys. **99**, 08Q703 (2006).
[7] J. C. Slonczewski, J. Magn. Magn. Mater. **159**, L1 (1996); J. Magn. Magn. Mater. **247**, 324 (2002)
[8] V. Kambersky, Can. J. Phys. **48**, 2906 (1970); V. Kambersky and C. E. Patton, Phys. Rev. B **11**, 2668 (1975).
[9] R. Duine, A. S. Nunez, J. Sinova, and A. H. MacDonald, arXiv:cond-mat/0703414.
[10] N. Smith, J. Appl. Phys. 90, 5768 (2001).
[11] S. E Barnes and S. Maekawa, Phys. Rev. Lett. **95**, 107204 (2005).